\documentclass[epsfig]{article}
\usepackage{amssymb}
\usepackage{graphicx}
\usepackage{amsmath}

\input epsf.sty
\input{tcilatex}
\begin{document}

\title{WMAP5 Observationnal Constraints on Braneworld New Inflation Model }
\author{R. Zarrouki$^{1}$, Z. Sakhi$^{2,3}$ and M. Bennai$^{1,3}$\thanks{%
E-mail adress: m.bennai@univh2m.ac.ma, bennai\_idrissi@yahoo.fr} \\
$^{\mathit{1}}${\small L.P.M.C,} {\small Facult\'{e} des Sciences Ben M'sik,
B.P. 7955, Universit\'{e} Hassan II-Mohammedia, Casablanca, Maroc. }\\
{\small \ }$^{\mathit{2}}${\small LISRI, Facult\'{e} des Sciences Ben M'Sik,
Universit\'{e} Hassan II-Mohammedia, Casablanca,\ Maroc, }\\
{\small \ }$^{3}${\small \ Groupement National de Physique des Hautes
Energies, Focal point, LabUFR-PHE, Rabat, Morocco.}}
\maketitle

\begin{abstract}
We study a new inflation potential in the framework of the \emph{%
Randall-Sundrum type 2} Braneworld model. Using the technic developped in%
\cite{Sanchez2007}, we consider both an monomial and a new inflation
potentials and apply the Slow-Roll approximation in high energy limit, to
derive analytical expression of relevant perturabtion spectrum. We show that
for some values of the parameter n of the potential $(V\left( \phi \right)
=V_{0}-\frac{1}{2}m^{2}\phi ^{2}+\frac{\alpha }{2n}\phi ^{2n})$\ we obtain
an perturbation spectrum wich present a good agreement with recent WMAP5
observations.

Keywords:\textbf{\ }\textit{RS Braneworld, }New inflation\textit{\
potential, Perturbation Spectrum, WMAP5.}

{\small PACS numbers: 98.80. Cq}
\end{abstract}

\date{}
\tableofcontents

\newpage

\newpage

\section{\protect\bigskip Introduction}

Recently Braneworld scenario\cite{braneworld1,braneworld2,braneworld3} has
become a central paradigm of modern inflationary cosmology. Standard
inflation has been mainly studied and was early confirmed by observations%
\cite{COBE}. Brane inflation is proposed to solve important cosmological
problems like as dark energy\cite{darkenergy}, tachyonic inflation\cite%
{taychons} or Black Holes systems\cite{BHRS1,BHRS2}. Others motivations are
observations on accelerating universe\cite{accelerating} as well as results
on interpretations of these phenomena in terms of scalar field dynamics.
Generally, scalar fields naturally arise in various particle physic theories
including string/M theory and expected to play a fundamental role in
inflation\cite{mtheory,scalar field}.

In \emph{Randall-Sundrum} model\cite{RSII} wich is one of the most studied
models, our four-dimensional universe is considered as a 3-brane embedded in
five-dimensional anti-de Sitter space-time($AdS5$), while gravity can be
propagated in the bulk. The most simplest inflationary models studied in the
context of \emph{Randall-Sundrum} scenario is the chaotic inflation\cite%
{Maartens}, but in relation with recents WMAP\textbf{\ }observations\cite%
{WMAP03,WMAP07,WMAP5}, more generalized models must be studied.

In this work, we are interested on a new inflationary model in the framework
of \emph{Randall-Sundrum} Braneworld inflation in relation with recent WMAP5%
\cite{WMAP5} for both monomial and New inflation potentials.

We first start in section 2, by recalling the foundations of the Braneworld
inflation precisely the modified \emph{Friedmann} eqs, and various
inflationary perturbation spectrum parameters. In the section 3 we present
our results for both Monomial and New inflation models. We have applied here
the Slow-roll approximation in the high energy limit to drive various
perturbations parameters spectrum for these models. We show that for some
values of the parameter n of the potential $(V\left( \phi \right) =V_{0}-%
\frac{1}{2}m^{2}\phi ^{2}+\frac{\alpha }{2n}\phi ^{2n}),$\ we obtain an
perturbation spectrum wich present a good agreement with recent WMAP5
observations. A conclusion and a perspective of this work are given in the
last section.

\section{Slow-roll Braneworld inflation}

\subsection{Randall-Sundrum model}

We start this section by recalling briefly some fundamentals of \emph{%
Randall-Sundrum} type II Braneworld model\cite{RSII}. In this model, our
universe is supposed living in a brane embedded in an Anti-de Sitter (AdS)
five-dimensional bulk spacetime. One of the most relevant consequences of
this model is the modification of the Friedmann equation for energy density
of the order of the brane tension, and also the appearance of an additional
term, usually considered as dark radiation term. \bigskip In the case where
the dark radiation term is neglected, the gravitationnal Einstein eqs, leads
to the modified \emph{Friedmann} equation on the brane as\cite{Maartens} 
\begin{equation}
H^{2}={\frac{8\pi }{3M_{pl}^{2}}}\rho \left[ 1+{\frac{\rho }{2\lambda }}%
\right]
\end{equation}%
with $\lambda $\ is the brane tension, $H$ is the \emph{Hubble} parameter
and $M_{pl}$ is th Planck mass$.$ It's clear that the crucial correction to
standard inflation is given by the density quadratic term $\rho ^{2}$. Brane
effect is then carried here by\ the deviation factor $\rho /2\lambda {,}$
with respect to unity. This deviation has the effect of modifying the
dynamics of the universe for density $\rho \gtrsim \lambda $. Note also that
in the limit $\lambda \rightarrow \infty ,$ we recover standard
four-dimensional standard inflation results. In inflationary theory, the
energy density $\rho ,$ and pressure $p,$ are expressed in term of inflaton
potential $V(\phi )$ as $\rho =\frac{1}{2}\dot{\phi}^{2}+V(\phi )$ and $p=%
\frac{1}{2}\dot{\phi}^{2}-V(\phi ),$ where $\phi $ is the inflaton field.

In inflation theory, the scalar potential $V(\phi )$, depending on the sclar
field $\phi ,$ play a fondamendal role and represent the initial vacum
energy responsible of inflation. Along with these equation, one also has a
second inflation \emph{Klein-Gordon} equation governing the dynamic of the
scalar field $\phi $%
\begin{equation}
\ddot{\phi}+3H\dot{\phi}+V^{\prime }(\phi )=0
\end{equation}%
This is a second order evolution equation which follows from conservation
condition of energy-momentum tensor $T_{\mu \nu }$. To calculate some
physical quantities as scale factor or perturbation spectrum, one has to
solve equations(1,2) for some specific potentials $V(\phi )$. To do so, the
Slow-Roll approximation was introduced and applied by many autors to drive
inflation perturbation spectrum\cite{liddle-2003}.

\subsection{Slow-Roll approximation and perturbation spectrum on brane}

Inflationary dynamics requires that inflaton field $\phi $ driving inflation
moves away from the false vacuum and slowly rolls down to the minimum of its
effective potential $V(\phi )$\cite{linde2005}. In this scenario, the
initial value $\phi _{i}=\phi \left( t_{i}\right) $ of the inflaton field
and the Hubble parameter $H$ are supposed large and the scale factor $%
a\left( t\right) $ of the universe growth rapidly. Applying the slow roll
approximation, $\dot{\phi}^{2}\ll V$ and $\ddot{\phi}\ll V^{\prime }$, to
brane field equations(1,2), we obtain: 
\begin{equation}
H^{2}\simeq {\frac{8\pi V}{3M_{4}^{2}}}\left( 1+{\frac{V}{2\lambda }}\right)
\,,\qquad \dot{\phi}\simeq -{\frac{V^{\prime }}{3H}}.
\end{equation}%
Note that slow roll approximation puts a constraint on the slope and the
curvature of the potential. This is clearly\ seen from the field expressions
of $\epsilon $ and $\eta $ parameters given by\cite{Maartens},\ 
\begin{eqnarray}
\epsilon &=&-\frac{\overset{\cdot }{H}}{H^{2}}\equiv {\frac{M_{4}^{2}}{4\pi }%
}\left( {\frac{V^{\prime }}{V}}\right) ^{2}\left[ \frac{\lambda (\lambda +V)%
}{(2\lambda +V)^{2}}\right] , \\
\eta &=&\frac{V^{\prime \prime }}{3H^{2}}\equiv {\frac{M_{4}^{2}}{4\pi }}%
\left( {\frac{V^{\prime \prime }}{V}}\right) \left[ \frac{\lambda }{2\lambda
+V}\right] .
\end{eqnarray}%
Slow-roll approximation takes place if these parameters are such that $%
\mathrm{max}\{\epsilon ,|\eta |\}\ll 1$ and inflationary phase ends when $%
\epsilon $ and $\left\vert \eta \right\vert $ are equal to one. Other
inflationary important quantity is the number $N_{e}$ of e-folding wich, in
slow roll approximation, reads as

\begin{equation}
N_{e}\simeq -{\frac{8\pi }{M_{4}^{2}}}\int_{\phi _{\mathrm{i}}}^{\phi _{%
\mathrm{f}}}{\frac{V}{V^{\prime }}}\left( 1+{\frac{V}{2\lambda }}\right)
d\phi .
\end{equation}%
where $\phi _{i}$ and $\phi _{f}$ stand for initial and final value of
inflaton.

Before proceeding, it is interesting to comment low and high energy limits
of these parameters. Note that at low energies where $V\ll \lambda $, the
slow-roll parameters take the standard form. At high energies $V\gg \lambda $%
, the extra contribution to the Hubble expansion dominates. The number of
e-folding in this case becomes $N_{e}\simeq -\frac{4\pi }{\lambda M_{4}^{2}}%
\int_{\phi _{i}}^{\phi _{f}}\frac{V^{2}}{V^{\prime }}d\phi .$

The inflationary spectrum perturbations is produced by quantum fluctuations
of fields around their homogeneous background values. Thus the scalar
amplitude $A_{\QTR{sc}{s}}^{2}$ of density perturbation, evaluated by
neglecting back-reaction due to metric fluctuation in fifth dimension, is
given by\cite{Maartens} 
\begin{equation}
A_{\QTR{sc}{s}}^{2}\simeq \left. \left( {\frac{512\pi }{75M_{4}^{6}}}\right) 
{\frac{V^{3}}{V^{\prime 2}}}\left[ {\frac{2\lambda +V}{2\lambda }}\right]
^{3}\right\vert _{k=aH}.
\end{equation}%
Note that for a given positive potential, the $A_{\QTR{sc}{s}}^{2}$
amplitude is increased in comparison with the standard result. In high
energy limit this quantity behaves as%
\begin{equation}
A_{S}^{2}\simeq \frac{64\pi }{75\lambda ^{3}M_{4}^{6}}\frac{V^{6}}{%
V^{^{\prime 2}}}.
\end{equation}%
On the other hand, using eqs(4,5), one can compute the perturbation
scale-dependence described by the spectral index $n_{\QTR{sc}{s}}\equiv
1+d\left( \ln A_{\QTR{sc}{s}}^{2}\right) /d\left( \ln k\right) $ and find%
\begin{equation}
n_{\QTR{sc}{s}}-1\simeq 2\eta -6\epsilon \,,
\end{equation}%
Note that at high energies $\lambda /V$, the slow-roll parameters are both
suppressed; and the spectral index is driven towards the Harrison-Zel'dovich
spectrum; $n_{\QTR{sc}{s}}\rightarrow 1$ as $V/\lambda \rightarrow \infty $.

In what follows, we shall apply the above Braneworld formalism by singling
out two specific kinds of inflaton potentials. These are the monomial and a
new inflation potentials recently studied by Boyanovsky et al.\cite%
{Sanchez2006} in standard inflation.

\section{Perturbation spectrum in Braneworld New inflation}

\qquad To begin, recall that chaotic inflationary model, which was first
introduced by Linde\cite{Linde82}, has been reconsidered recently by several
authors in the context of Braneworld scenario\cite{Maartens,Paul}. In the
present work, we are interested by others types of potential inflation. The
autors of \cite{Sanchez2006}\ have shwon that combining the WMAP data with
the slow roll expansion constraints the inflaton potential to have the form 
\begin{equation}
V(\phi )=N_{e}M^{4}w\left( \chi \right)
\end{equation}%
$M$ is the inflation energy scale determined by the amplitude of the scalar
adiabatic fluctuations\cite{M} to be $M\sim 0.00319$ $M_{Pl}=0.77\times
10^{16}GeV$, where a dimensionless rescaled field variable is introduced%
\begin{equation}
\chi =\frac{\phi }{\sqrt{N_{e}}M_{pl}}.
\end{equation}%
here $N_{e}$ is the number of efolds.

In this new notation, the Slow-Roll parameters become 
\begin{eqnarray}
\epsilon &=&\frac{\lambda }{4\pi N_{e}^{2}M^{4}}\left( \frac{w^{\prime
}\left( \chi \right) ^{2}}{w\left( \chi \right) ^{3}}\right) \\
\eta &=&\frac{\lambda }{4\pi N_{e}^{2}M^{4}}\left( \frac{w^{\prime \prime
}\left( \chi \right) }{w\left( \chi \right) ^{2}}\right)
\end{eqnarray}%
where the prime stands for derivative with respect to $\chi :w^{\prime }=%
\frac{dw}{d\chi }$ and $w^{\prime \prime }=\frac{d^{2}w}{d\chi ^{2}}$.

The perturbations parameters are now expressed in term of the new variable
as 
\begin{equation}
N_{e}=\frac{1}{-\frac{4\pi }{\lambda }M^{4}\int_{\chi _{c}}^{\chi _{end}}%
\frac{w\left( \chi \right) ^{2}}{w^{\prime }\left( \chi \right) }d\chi }
\end{equation}%
where $\chi _{c}$ is the value of $\chi $ corresponding to $N_{e}$ e-folds
before the end of inflation, and $\chi _{end}$ is the value of $\chi $ at
the end of inflation.

Other perturbation quantities are also calculated in term of $w\left( \chi
\right) $ as 
\begin{equation}
A_{\QTR{sc}{s}}^{2}=\frac{64\pi N_{e}^{5}M^{16}}{75\lambda ^{3}M_{pl}^{4}}%
\left( \frac{w\left( \chi \right) ^{6}}{w^{\prime }\left( \chi \right) ^{2}}%
\right) ,
\end{equation}%
The spectral and running index are now respectively written in the following
form 
\begin{eqnarray}
n_{s}-1 &=&\frac{\lambda }{4\pi N_{e}^{2}M^{4}}\left( 2\frac{w^{\prime
\prime }\left( \chi \right) }{w\left( \chi \right) ^{2}}-6\frac{w^{\prime
}\left( \chi \right) ^{2}}{w\left( \chi \right) ^{3}}\right) , \\
\frac{dn_{s}}{d\ln k} &=&-\frac{\lambda ^{2}}{8\pi ^{2}M^{8}N_{e}^{4}}\left(
-\frac{8w^{\prime \prime }\left( \chi \right) w^{\prime }\left( \chi \right)
^{2}}{w\left( \chi \right) ^{5}}+9\frac{w^{\prime }\left( \chi \right) ^{4}}{%
w\left( \chi \right) ^{6}}+\frac{w^{\prime }\left( \chi \right) w^{\prime
\prime \prime }\left( \chi \right) }{w\left( \chi \right) ^{4}}\right)
\end{eqnarray}%
\ Finally, the ratio of tensor to scalar perturbations $r$ reads 
\begin{equation}
r=\frac{6\lambda }{\pi N_{e}^{2}M^{4}}\left( \frac{w^{\prime }\left( \chi
\right) ^{2}}{w\left( \chi \right) ^{3}}\right) .
\end{equation}%
In what follows we will determine all this inflationary perturbation
spectrum paramters at $\chi =\chi _{c}$, for a monomial and a new inflation
potential and compare our results with recent $WMAP$ experimental data in
the later case.

\subsection{Monomial potential}

Let as begin by an monomial potential which generalize the chaotic one$.$
Chaotic inflation was mainly studied in the context of standard\cite{Linde82}%
, Brane\cite{Maartens,Paul} and recently Chaplygin inflation on the brane 
\cite{Chaplygin inflation}. Here, we consider a general potantial of the
form 
\begin{equation}
V\left( \phi \right) =\frac{\alpha }{2n}\phi ^{2n},
\end{equation}%
where $\alpha $ and $n$ are constants.

In term of \ $\chi ,$ we get 
\begin{equation}
w\left( \chi \right) =\frac{\chi ^{2n}}{2n}
\end{equation}%
where we have used 
\begin{equation}
M^{4}=\alpha N_{e}^{n-1}M_{pl}^{2n}
\end{equation}%
Thus, slow-roll parameters are represented by 
\begin{eqnarray}
\varepsilon &=&\frac{2\lambda n^{3}}{\pi N_{e}^{2}M^{4}}\left( \frac{1}{\chi
_{c}^{2n+2}}\right) \\
\eta &=&\frac{\lambda n^{2}\left( 2n-1\right) }{\pi N_{e}^{2}M^{4}}\left( 
\frac{1}{\chi _{c}^{2n+2}}\right)
\end{eqnarray}%
and scalar spectral index $n$ and the ratio $r$ are respectively expressed as

\begin{eqnarray}
n_{s}-1 &=&-\frac{2\lambda n^{2}}{\pi N_{e}^{2}M^{4}}\left( 4n+1\right)
\left( \frac{1}{\chi _{c}^{2n+2}}\right) \\
r &=&48\frac{\lambda n^{3}}{\pi N_{e}^{2}M^{4}}\left( \frac{1}{\chi
_{c}^{2n+2}}\right)
\end{eqnarray}%
Finally the running index takes the following expression 
\begin{equation}
\frac{dn_{s}}{d\ln k}=-\frac{4\lambda ^{2}}{\pi ^{2}M^{8}N_{e}^{4}\chi
_{c}^{4n+4}}\left( 4n^{6}+5n^{5}+n^{4}\right)
\end{equation}%
Inflation ends at $\chi _{_{end}}$ $=0$, thus the value of the dimensionless
field $\chi _{c}$ before the end of inflation is 
\begin{equation}
\chi _{c}^{2n+2}=\frac{\lambda n^{2}\left( 2n+2\right) }{\pi N_{e}M^{4}}
\end{equation}%
Winding this result, various perturbations parameters are obtained in term
of potential parameter n and e-fold number $N_{e}$ 
\begin{eqnarray}
\varepsilon &=&\frac{n}{N_{e}\left( n+1\right) }\text{ },\text{\ \ \ \ \ \ \
\ \ \ \ }\eta =\text{\ }\frac{2n-1}{2N_{e}\left( n+1\right) },\text{\ \ \ \ }
\\
n_{s}-1 &=&-\frac{4n+1}{N_{e}\left( n+1\right) },\text{ \ \ \ \ \ \ \ \ \ \ }%
r=24\frac{n}{N_{e}\left( n+1\right) }\text{ \ },\text{ \ } \\
\text{\ }\frac{dn_{s}}{d\ln k} &=&-\frac{\left( 4n^{2}+5n+1\right) }{%
N_{e}^{2}\left( n+1\right) ^{2}}
\end{eqnarray}

It will be interessing to study the variation of these perturbation
parameters as function of potential parameter n, and compare the results to
recent WMAP5 observations. In the following, we do this for a more
generalized new inflation potential

\subsection{New inflation model}

Consider now a new inflation potential of the form\cite{Sanchez2007}

\begin{equation}
w\left( \chi \right) =w_{0}-\frac{1}{2}\chi ^{2}+\frac{g}{2n}\chi ^{2n}\text{%
\ }
\end{equation}%
where $w_{0}$ and the coupling $g$ are dimensionless.

In ref.\cite{Sanchez2007} the authors used this model in standard inflation
and they have shown that for lower values of $n$ the results reproduce
observation. In the present work, we reproduce a new results for all known
inflation spectrum\ parameters, but in the context of \emph{Randall-Sundrum}
Braneworld model.

New inflation model described by the dimensionless potential given by eq.$%
(31)$ have a minimum at $\chi _{_{0}}$ which is the solution to the
following conditions 
\begin{equation}
w^{\prime }\left( \chi _{0}\right) =w\left( \chi _{0}\right) =0
\end{equation}%
These conditions yield 
\begin{equation}
g=\frac{1}{\chi _{0}^{2n-2}}\text{ \ },\text{ \ \ \ \ \ \ \ \ \ }w_{0}=\frac{%
\chi _{0}^{2}}{2n}\left( n-1\right) 
\end{equation}%
As using the previous results the equation $(31)$ becomes 
\begin{equation}
w\left( \chi \right) =\frac{\left( n-1\right) }{2n}\chi _{0}^{2}-\frac{\chi
^{2}}{2}+\frac{\chi _{0}^{2-2n}}{2n}\chi ^{2n}
\end{equation}%
$\chi _{0}$ determines the scale of symmetry breaking $\phi _{0}$ of the
inflaton potential upon the rescaling eq.$(11)$, namely 
\begin{equation}
\phi _{0}=\sqrt{N_{e}}M_{pl}\chi _{0}
\end{equation}%
It is convenient to introduce the dimensionless variable 
\begin{equation}
x=\frac{\chi }{\chi _{0}}
\end{equation}%
Then, from eq.$(36)$, the potential of inflation model eq.$(34)$ takes the
form 
\begin{equation}
w\left( x\right) =\ \frac{\chi _{0}^{2}}{2n}\left[ n\left( 1-x^{2}\right)
+x^{2n}-1\right] ,\text{ \ \ \ \ \ \ \ \ \ \ \ \ broken symmetry}
\end{equation}%
Inflation ends when the inflaton field arrives at the minimum of the
potential. As shown by \emph{Linde}\cite{linde2005}, the symmetry is browken
for non vanishing minimun of the potential. Thus, for our new inflation
model eq.$(34)$ inflation ends for 
\begin{equation}
\chi _{end}=\chi _{0}
\end{equation}%
According to the new varaible $x,$ the condition eq.$(14)$ becomes 
\begin{equation}
1=\frac{\pi N_{e}M^{4}\chi _{0}^{4}}{\lambda n^{2}}I_{n}\left( X\right) 
\end{equation}%
where 
\begin{equation}
I_{n}\left( X\right) =\int_{X}^{1}\frac{\left( n\left( 1-x^{2}\right)
+x^{2n}-1\right) ^{2}}{\left( 1-x^{2n-2}\right) }\frac{dx}{x}
\end{equation}%
and \ \ \ \ \ \ \ \ \ \ \ \ \ \ \ \ \ \ \ \ \ \ \ \ \ \ \ \ \ \ \ \ \ \ \ \
\ \ \ \ \ \ \ \ 
\begin{equation}
\ \ X=\frac{\chi _{c}}{\chi _{0}}
\end{equation}%
For small field and $X\longrightarrow 1^{-\text{ }}$ the integral $I_{n}(X)$
obviously vanishes and by expanding the potential (eq.34) near the minimum $%
\chi _{0}$ $\left( n\succ 1\right) $ 
\begin{equation}
w\left( \chi \right) \sim \frac{\left( 2n-2\right) \left( \chi -\chi
_{0}\right) ^{2}}{2}
\end{equation}%
The expression approached of the potential (eq.42) allows us to recover the
expression of the monomial potential (eq.20) for $n=1$ by simple shift 
\begin{equation}
\chi \longrightarrow \sqrt{\left( 2n-2\right) }\left( \chi -\chi _{0}\right)
;\text{\ \ \ \ \ \ \ \ \ }\left( n\succ 1\right) 
\end{equation}%
So, we can determine all inflationary perturbation spectrum paramters near
the minimum $X=1.$ Therefore for $X\sim 1$ the quadratic monomial is an
excellent approximation to the family of higher degree potentials.

The slow-roll parameters become in term of the varaible X 
\begin{eqnarray}
\varepsilon &=&\frac{2nI_{n}\left( X\right) }{N_{e}}\frac{\left(
-X+X^{2n-1}\right) ^{2}}{\left( n\left( 1-X^{2}\right) +X^{2n}-1\right) ^{3}}%
\text{ \ } \\
\text{ \ \ }\eta &=&\frac{I_{n}\left( X\right) }{N_{e}}\frac{\left(
-1+\left( 2n-1\right) X^{2n-2}\right) }{\left( n\left( 1-X^{2}\right)
+X^{2n}-1\right) ^{2}}
\end{eqnarray}%
In the following, we study the variation of these parameters as function of
X, by numerical calculations for $N_{e}=50.$

\FRAME{dtbpFU}{4.0707in}{3.0217in}{0pt}{\Qcb{Figure 1: $\protect\varepsilon $
$vs$ $X$ for new inflation potential ($n=2,3,4).$}}{}{kbs15e00.wmf}{\special%
{language "Scientific Word";type "GRAPHIC";maintain-aspect-ratio
TRUE;display "USEDEF";valid_file "F";width 4.0707in;height 3.0217in;depth
0pt;original-width 4.491in;original-height 3.3261in;cropleft "0";croptop
"1";cropright "1";cropbottom "0";filename 'KBS15E00.wmf';file-properties
"XNPEU";}}\FRAME{dtbpFU}{3.8994in}{3.0217in}{0pt}{\Qcb{Figure 2: $\protect%
\eta $ $vs$ $X$ for new inflation potentialor ($n=2,3,4).$}}{}{Figure}{%
\special{language "Scientific Word";type "GRAPHIC";maintain-aspect-ratio
TRUE;display "USEDEF";valid_file "T";width 3.8994in;height 3.0217in;depth
0pt;original-width 4.3016in;original-height 3.3261in;cropleft "0";croptop
"1";cropright "1";cropbottom "0";tempfilename
'KLMX2201.wmf';tempfile-properties "XPR";}}The figures 1 and 2 show the
increasing behavior of the two functions $\varepsilon $ and $\eta $ for
small values for $X$ and for large $X$ , the both functions become constant.
We can remark again that a large domain of variation of $X$ verifies the
conditions of inflation since during inflation we have $\epsilon \ll 1$ and $%
\mid \eta \mid \ll 1.$

\bigskip On the other hand, observations combining WMAP5, BAO$\left( \text{%
Baryon Acoustic Oscillations}\right) $ and SN$\left( \text{Type Ia supernovae%
}\right) $ data\cite{WMAP5}, yields

\begin{equation}
n_{s}=0.960_{\text{ }-0.013}^{\text{ }+0.014}\text{ \ }(95\%\text{ CL})\text{
}
\end{equation}

\begin{equation}
r<0.20\text{ }(95\%\text{ CL})\text{ \ \ \ \ \ \ \ \ \ \ \ }
\end{equation}

\begin{equation}
-0.0728<\frac{dn_{s}}{dlnk}<0.0087\text{\ \ \ \ }(95\%\text{ CL})\text{ \ \ }%
\ 
\end{equation}%
In our variable X, the spectral index becomes%
\begin{equation}
n_{s}-1=\frac{2I_{n}\left( X\right) }{N_{e}\left( n\left( 1-X^{2}\right)
+X^{2n}-1\right) ^{2}}\left[ -6n\frac{\left( -X+X^{2n-1}\right) ^{2}}{\left(
n\left( 1-X^{2}\right) +X^{2n}-1\right) }-1+\left( 2n-1\right) X^{2n-2}%
\right]
\end{equation}

In figue 3, we plot this parameter as function of X. So, we can remark that
the window of consistency with the WMAP5 + BAO+SN data narrows for growing $%
n.$ Thus we obtain a good potential expression for small $n<3$ and large X.%
\FRAME{dtbpFU}{4.0819in}{3.0139in}{0pt}{\Qcb{Figure 3: $n_{s}$ $vs$ $X$ for
new inflation potential($n=2,3,4)$. Horizontal doted line corresponds to the
lower limit $n_{s}=0.9392$ ($95\%CL$) from WMAP5,BAO and SN}}{}{Figure}{%
\special{language "Scientific Word";type "GRAPHIC";maintain-aspect-ratio
TRUE;display "USEDEF";valid_file "T";width 4.0819in;height 3.0139in;depth
0pt;original-width 4.5039in;original-height 3.3174in;cropleft "0";croptop
"1";cropright "1";cropbottom "0";tempfilename
'KLMX2302.wmf';tempfile-properties "XPR";}} Other spectrum parameter is the
ratio $r$ presented by%
\begin{equation}
r=48\frac{nI_{n}\left( X\right) }{N_{e}}\frac{\left( -X+X^{2n-1}\right) ^{2}%
}{\left( n\left( 1-X^{2}\right) +X^{2n}-1\right) ^{3}}
\end{equation}%
\FRAME{dtbpFU}{3.9574in}{3.0208in}{0pt}{\Qcb{Figure 4: $r$ $vs$ $X$ for new
inflation potential ($n=2,3,4)$. Horizontal doted line corresponds to the
upper limit $r=0.2$ ($95\%CL$) from WMAP5,BAO and SN}}{}{Figure}{\special%
{language "Scientific Word";type "GRAPHIC";maintain-aspect-ratio
TRUE;display "USEDEF";valid_file "T";width 3.9574in;height 3.0208in;depth
0pt;original-width 4.3673in;original-height 3.3261in;cropleft "0";croptop
"1";cropright "1";cropbottom "0";tempfilename
'KLMX2303.wmf';tempfile-properties "XPR";}}This figure shows the same
behavior as figure1 since $r=24\epsilon .$ Note that the observational
result is reproduced for small values of $X$ where the three curves are
almost confounded. \FRAME{dtbpFU}{3.8277in}{3.0364in}{0pt}{\Qcb{Figure 5: $r$
$vs$ $n_{s}$ for new inflation potential($n=2,3,4).$}}{}{Figure}{\special%
{language "Scientific Word";type "GRAPHIC";maintain-aspect-ratio
TRUE;display "USEDEF";valid_file "T";width 3.8277in;height 3.0364in;depth
0pt;original-width 4.222in;original-height 3.3425in;cropleft "0";croptop
"1";cropright "1";cropbottom "0";tempfilename
'KLMX2304.wmf';tempfile-properties "XPR";}}To confronte simutaniously the
observales r and n$_{s}$ with observation, it will be interesting to study
the relative variation of these parameters. In figure 5 we have ploted $r$ $%
vs$ $n_{s}.$ The black dot corresponds to chaotic potential $\frac{\alpha
\phi ^{2}}{2}\left( X=1,\text{ }n_{s}=0.95\text{ and }r=0.24\right) $.
Values below the black dot corresponds to $X\prec 1$ which constitutes the
region where the observation results are recovered. For values corresponding
to $X\succ 1$, the parameter values become in disagreement with observation
notably for $r.$

We have also calculated the running index $\frac{dn_{s}}{d\ln k}$ wich is
given by

\begin{equation}
\frac{dn_{s}}{d\ln k}=-\frac{I_{n}\left( X\right) ^{2}}{8n^{4}N_{e}^{2}}%
\left( -\frac{8v^{\prime \prime }\left( X\right) v^{\prime }\left( X\right)
^{2}}{v\left( X\right) ^{5}}+9\frac{v^{\prime }\left( X\right) ^{4}}{v\left(
X\right) ^{6}}+\frac{v^{\prime }\left( X\right) v^{\prime \prime \prime
}\left( X\right) }{v\left( X\right) ^{4}}\right)
\end{equation}%
where%
\begin{eqnarray}
v\left( X\right) &=&\frac{\left[ n\left( 1-X^{2}\right) +X^{2n}-1\right] }{2n%
} \\
v^{\prime }\left( X\right) &=&\frac{\partial v\left( X\right) }{\partial X} 
\notag
\end{eqnarray}%
We observe, in figure 6, that for any values of $X$ the experimental data is
verified. Thus all members of new inflation potential family predict a small
and negative running\FRAME{dtbpFU}{4.0171in}{3.0217in}{0pt}{\Qcb{Figure 6: $%
\frac{dn_{s}}{d\ln k}$ $\left( \text{Running of the scalar index}\right) vs$ 
$X$ for $n=2,3,4$ for new inflation potential}}{}{Figure}{\special{language
"Scientific Word";type "GRAPHIC";display "USEDEF";valid_file "T";width
4.0171in;height 3.0217in;depth 0pt;original-width 4.433in;original-height
3.3261in;cropleft "0";croptop "1";cropright "1";cropbottom "0";tempfilename
'KLMX2405.wmf';tempfile-properties "XPR";}}

For the some reasons as above, we plot in the last figure, the variation of $%
\frac{dn_{s}}{d\ln k}$ $vs$ $n_{s}.$ \FRAME{dtbpFU}{4.0914in}{3.1272in}{0pt}{%
\Qcb{Figure 7: $\frac{dn_{s}}{d\ln k}$ $vs$ $n_{s}$ for new inflation
potential ($n=2,3,4).$}}{}{Figure}{\special{language "Scientific Word";type
"GRAPHIC";maintain-aspect-ratio TRUE;display "USEDEF";valid_file "T";width
4.0914in;height 3.1272in;depth 0pt;original-width 4.5152in;original-height
3.4445in;cropleft "0";croptop "1";cropright "1";cropbottom "0";tempfilename
'KLMX2406.wmf';tempfile-properties "XPR";}}We note that for $n_{s}$
experimental data ($0.9392\prec n_{s}\prec 0.9986$), all values of $\frac{%
dn_{s}}{d\ln k}$ are conforme with observation for any n. \ As before, the
black dot corresponds to chaotic potential $\frac{\alpha \phi ^{2}}{2}\left(
X=1,\text{ }n_{s}=0.95\text{ and }\frac{dn_{s}}{d\ln k}=-0.0010\right) $.
Thus, a good values of $X$ wich are in agreement within obervation
correspond to $X\prec 1.$

\section{Conclusion}

In this work, we have studied a new inflation potential in the framework of
Braneworld \emph{Rundall-Sundrum} type 2 model. We have applied Slow-Roll
approximation in high energy limit in order to derive analytical expressions
for various perturbation spectrum (n$_{s}$, r and $\frac{dn_{s}}{d\ln k})$.
We have considered a monomial and a new inflation\ potentials to study the
behaviours of inflation spectrum for various values of n. We have shown that
for some values of the parameter n of the potential $(V\left( \phi \right)
=V_{0}-\frac{1}{2}m^{2}\phi ^{2}+\frac{\alpha }{2n}\phi ^{2n})$\ our results
are in a good agreement with recent WMAP5 observations, specially for small
fields.

\end{document}